\documentclass[11pt,dvips]{article}
\usepackage{epsfig,times}
\makeatletter
\renewcommand{\section}{\@startsection{section}{1}{0in}
	{0.4\baselineskip}{0.1\baselineskip}{\Large\bf}}
\renewcommand{\subsection}{\@startsection{subsection}{2}{0in}
	{0.25\baselineskip}{-\baselineskip}{\large\bf}}
\renewcommand{\subsubsection}{\@startsection{subsubsection}{3}{0in}
	{0.1\baselineskip}{-\baselineskip}{\normalsize\bf}}
\makeatother
\begin{document}

%
%
{\it OG 3.2.12}

\begin{center}
%

{\LARGE \bf On Nonlinear Alfv\'en Waves Generated by Cosmic Ray Streaming Instability}
\footnote {Published in Proc. 26th ICRC 4, 233, Salt Lake City, USA, 1999}
\end{center}

\begin{center}
%
%
{\bf V.N.Zirakashvili$^1$, V.S.Ptuskin$^1$, and H.J.V\"olk$^2$}\\
{\it $^1$Institute for Terrestrial Magnetism, Ionosphere and Radiowave
Propagation,Russian Academy of Sciences (IZMIRAN), 142190, Troitsk, Moscow Region, Russia\\
$^2$Max-Planck-Institut f\"ur Kernphysik,Postfach 103980, D-69029
Heidelberg,Germany}
\end{center}

\begin{center}
{\large \bf Abstract\\}
\end{center}
\vspace{-0.5ex}
%
%
Nonlinear damping of parallel propagating Alfv\'en waves in high-$\beta$ plasma is considered. Trapping of thermal ions and Coulomb collisions are taken into account. Saturated damping rate is calculated. Applications are made for cosmic ray propagation in the Galaxy.
%

\vspace{1ex}

%
%

\section{Introduction.}
It is well known that the cosmic ray streaming instability can play an important role in
processes of diffusive shock acceleration and galactic propagation of cosmic ray particles since it can supply Alfv\'en waves that scatter the particles on pitch angle (Lerche,1967, Kulsrud \& Pearce,1969, Wentzel,1969). In order to balance wave generation some damping mechanism is usually considered. As Alfv\'en waves are weakly linearly damped, various
nonlinear effects are currently used. Cosmic ray streaming generates waves
in one hemisphere of wave vectors. It is well known that such waves are not
subject to any damping in incompressible magneto-hydrodynamics. Using compressibility results in a pondermotive force that gives a second order plasma velocity and electric field perturbations along the mean magnetic field. This perturbations can yield wave steepening as well as nonlinear damping, if kinetic
effects of thermal particles are included. Those effects were taken into
account in order to obtain nonlinear damping rates  of parallel propagating
Alfv\'en waves (Lee \& V\"olk, 1973, Kulsrud, 1978, Achterberg, 1981, Fedorenko, 1992). The importance of trapping of thermal particles for nonlinear
dissipation of sufficiently strong waves that results in saturation of wave damping was also understood many years ago (Kulsrud, 1978, V\"{o}lk \& Cesarsky 1982, Fedorenko, 1992). Corresponding saturated damping rates that take into account dispersive effects were calculated.\\
Nevertheless dispersive effects can be rather small for Alfv\'en waves  that are in resonance with galactic cosmic ray nuclei. Hence the effect of Coulomb collisions can be important. In this paper we will derive the nonlinear Alfv\'en wave damping rate in presence of thermal collisions.
\section{Basic equations.}
We will consider Alfv\'en waves propagating in one direction along the ambient
magnetic field. It is convenient to write the equations in the frame moving with the waves. In such a frame there are only quasistatic magnetic and electric fields  slowly varying in time due to wave dispersion and nonlinear effects. The case of a high-$\beta $ Maxwellian plasma will be considered below. Electric fields are negligible for nonlinear damping in such a plasma. We will investigate
waves with wavelengths much greater thermal particles gyroradii and will use
drift equations for distribution function of those particles (Chandrasekhar,1960).
\begin{equation}
\frac{\partial F}{\partial t}+v\mu ({\bf b}\nabla )F+\frac{
1-\mu ^{2}}{2}v\frac{\partial F}{\partial \mu }\nabla {\bf{b}
}=St~F  \label{1}
\end{equation}
Here $F$ is the velocity distribution of thermal particles that
is averaged on gyroperiod, $v$ is particle velocity, {\bf b}={\bf B}/B is the unit vector along the magnetic
field {\bf B}, $\mu =${\bf pB}/B is the cosine of the pitch angle of the
particle. The right hand side of equation (1) describes collisions of particles.
For Maxwell equations it is necessary to know the flux of particles. It
is given by drift theory (Chandrasekhar,1960):
\begin{equation}
{\bf{J}}_{\perp }=\frac{1-\mu ^{2}}{2}\frac{v^{2}}{\Omega }
{\bf{b}}\times \left[ \nabla F+\mu \frac{\partial F}{\partial \mu }(
{\bf{b}}\nabla ){\bf{b}}\right]   \label{2}
\end{equation}
Here $\Omega $ is particle gyrofrequency in local field.
The last term on the left hand side of Eq.(1) describes mirroring of
particles. As in this frame the field is static, the particle energy is constant, and in a time asymptotic state wave dissipation is absent without collisions. In the presence of wave excitation we shall only deal with the time asymptotic state in the following. We shall use for the collision
operator a simplified form $St~F=\Delta _{v}\nu $v$^{2}\left(
F-F_{M}\right) $, where $F_{M}$ is the Maxwellian distribution function shifted by the Alfv\'en velocity v$_{a}$, $\Delta _{v}$ is the Laplace operator in velocity space and $\nu $
is the collision frequency. This operator tends to make the particle distribution function Maxwellian. Introducing the coordinate $s$ along the magnetic field, and the distribution function $f=F-F_{M}$ one
obtains the following equation for $f$:
\begin{equation}
v\mu \frac{\partial f}{
\partial s}-\frac{1-\mu ^{2}}{2}v\frac{\partial f}{\partial
\mu }\frac{\partial \ln B(s)}{\partial s}-\Delta _{v}\nu v
^{2}f=\frac{1-\mu ^{2}}{2}v\frac{\partial F_{M}}{\partial \mu
}\frac{\partial \ln B(s)}{\partial s}  \label{3}
\end{equation}
For sufficiently small magnetic field perturbations (conditions for that
case will be derived later) one can neglect the mirroring term on the left
hand side of Eq.(3). Without collisions this leads to the well known nonlinear damping mentioned above. We shall take into account the mirroring term here. We will use standard quasilinear theory (Galeev \& Sagdeev,1979). The function $f$ can be written in the form $f=f_o+\delta f$, where $f_o=\langle f\rangle $ is ensemble averaged distribution function $f$. We are interested in the case of a small magnetic field amplitude $A<<1$, where ${\bf A}=({\bf B}-{\bf B}_o)/B_o$. Taking also into account that mirroring is sufficient for small $\mu <<1$ particles we leave in the collision operator the second derivative on $\mu $ only and come to the equation:
\begin{equation}
v\mu \frac {\partial f}{\partial s}-\frac v4\frac \partial {\partial \mu }(f+F_M)\frac {\partial A^2(s)}{\partial s}-\nu \frac {\partial ^2f}{\partial \mu ^2}=0 \label {4}
\end{equation}
Taking into account that average distribution function is $s$ independent one can obtain equation for Fourier transform $\delta f_k=\int ds\delta f(s)\exp (-isk)$:
\begin{equation}
ikv\mu \delta f_k-\nu \frac {\partial ^2\delta f_k}{\partial \mu ^2}=\frac 14ikvA^2_k\frac \partial {\partial \mu}(F_M+f_o) \label{5}
\end{equation}
The functions $f_o$ and $\delta f_k$ are peaked near $\mu=0$. It is convenient to introduce Fourier transform on $\mu $ $\tilde{f}_o(\xi )=\int d\mu f_o(\mu )\exp (-i\xi \mu )$ and $\delta \tilde{f}_k(\xi )=\int d\mu \delta f_k(\mu )\exp (-i\xi \mu )$. Then Eq. (5) will cast
\begin{equation}
kv\frac {\partial \delta \tilde{f}_k}{\partial \xi} +\nu \xi ^2\delta \tilde{f}_k=\frac v4ikA^2_k\left( 2\pi \delta (\xi )\frac {\partial F_M}{\partial \mu}|_{\mu =0}+i\xi \tilde{f}_o(\xi )\right) \label{6}
\end{equation}
This equation has a solution
\begin{equation}
\delta \tilde{f}_k=i\frac k{4|k|}A^2_k\int \limits ^{+\infty}_{-\infty }d\xi '\theta(k(\xi -\xi '))\left( 2\pi \delta (\xi ')\frac {\partial F_M}{\partial \mu}|_{\mu =0}+i\xi '\tilde{f}_o(\xi ')\right)\exp \left( -\frac \nu {3kv}(\xi ^3-\xi '^3)\right) \label {7}
\end{equation}
After ensemble averaging of Eq.(4) and using expression (7) one obtains an equation for $\tilde{f}_o(\xi )$:
$$
\nu \xi ^2\tilde{f}_o(\xi )-\frac {i\xi }{16}\int dkdk_1\int \limits ^{+\infty }_{-\infty}d\xi '|k|v\theta (k(\xi -\xi '))I(k_1)I(k+k_1)
\cdot $$
\begin{equation}
\cdot
\left( 2\pi \delta (\xi ')\frac {\partial F_M}{\partial \mu }|_{\mu =0}+i\xi '\tilde{f}_o(\xi ')\right) \exp \left( -\frac \nu {3kv}(\xi ^3-\xi '^3)\right)=0 \label{8}
\end{equation}
Here $I(k)$ is spectrum of Alfv\'en waves normalized to magnetic energy of the mean field: $\left\langle \delta B^2\right\rangle =B^2_o\int dkI(k)$. Wavenumbers with $+(-)$ sign correspond to right(left) hand circular polarized wave. Equation obtained describes influence of waves on the mean distribution function of thermal particles, in particular, well known in plasma theory quasilinear ``plateau'' formation breaking by thermal collisions (Galeev \& Sagdeev,1979). The solution of this equation should be substituted into expression (7). This expression, together with the expression for the flux (2) determines the nonlinear electric current density (the input of thermal protons is taken into account only)
\begin{equation}
{\bf J}_k=i\pi k\int \limits ^{+\infty }_{-\infty }dk_1\int \limits ^\infty _0v^2dvMc\frac {v^2}{B_o}[{\bf A}_{k-k_1}{\bf \times e}_z]\delta \tilde{f}_{k_1}|_{\xi =0}. \label {9}
\end{equation}
Substituting this current into Maxwell equations and ensemble averaging one can derive an equation for the Alfv\'en wave spectrum
$dI(k)/dt=-2\Gamma_{NL}I(k)$
with the nonlinear Alfv\'en wave damping rate:
$$
\Gamma _{NL}=-\frac \pi 8\frac {Mv_a}{B^2_o}k\int \limits ^{+\infty }_{-\infty }dk_1 I(-k_1)\int \limits ^\infty _02\pi v^2dvv^2\int \limits ^{+\infty}_{-\infty}d\xi '\frac {k+k_1}{|k+k_1|}\theta (-\xi '(k+k_1))\cdot
$$
\begin{equation}
\cdot \left( 2\pi \delta (\xi ')\frac {\partial F_M}{\partial \mu}|_{\mu =0}+i\xi '\tilde{f}_o(\xi ')\right) \exp \left( \frac {\nu \xi '^3}{3(k+k_1)v}\right) \label {10}
\end{equation}
where $M$ is the ion mass and
\begin{equation}
\frac {\partial F_M}{\partial \mu }|_{\mu =0}=-\frac{nv_av}{(2\pi )^{3/2}v%
_{T}^5}\exp \left[ -\frac{v^2+v_a^2}{2v_{T}^2}\right] \label{11}
\end{equation}
Here $n$ is the plasma density and $v_{T}$ is the thermal velocity.\\
It is useful to transform Eq.(8) to a form more convenient for applications. It is possible to invert the integral operator and obtain the following equation:
\begin{equation}
8\xi ^2\int \limits ^{+\infty}_{-\infty}d\xi '\xi '^3\tilde{f}_o(\xi ')\int \limits ^{+\infty}_{-\infty}\frac {d\eta }{2\pi }\frac{\exp \left( \frac {i\eta }3(\xi ^3-\xi '^3)\right)}{\int \limits ^{+\infty}_{-\infty}dkdk_1\frac {I(k_1)I(k+k_1)k^2v^2}{\nu ^2+\eta ^2k^2v^2}}+
\xi \tilde{f}_o(\xi )=2\pi i\delta(\xi )\frac {\partial F_M}{\partial \mu}|_{\mu =0} \label{12}
\end{equation}
One should solve Eq.(12) in order to use expression (10) except in the case when the collision frequency is large enough and a ``plateau'' is absent. In this case one can neglect $\tilde{f}_o$ in expression (10) and obtain the well known unsaturated nonlinear damping rate (Lee \& V\"olk, 1973, Kulsrud, 1978, Achterberg, 1981, Fedorenko, 1992):
\begin{equation}
\Gamma ^{(0)}_{NL}=\frac 18\sqrt {2\pi }v_Tk\int \limits ^{+\infty}_{-\infty}dk_1I(k_1)\frac {k-k_1}{|k-k_1|}\label {13}
\end{equation}
In the opposite case of small $\nu $ one should use Eq. (12) and put $\nu =0$:
\begin{equation}
\frac \partial {\partial \xi}\frac 1{\xi ^2}\frac \partial {\partial \xi}\xi \tilde {f}_o-\frac 18\left\langle A^2\right\rangle ^2\xi \tilde {f}_o=-\frac {i\pi }4\delta (\xi )\left\langle A^2\right\rangle ^2 \frac {\partial F_M}{\partial \mu}|_{\mu =0}\label {14}
\end{equation}
Here $\left\langle A^2\right\rangle =\int \limits ^{+\infty}_{-\infty}dkI(k)$. Substituting the solution of this equation into expression (10) and expanding the exponent one can obtain the saturated damping rate:
\begin{equation}
\Gamma ^{sat}_{NL}=\nu _o\left\langle A^2\right\rangle ^{-3/2}k\int \limits ^{+\infty}_{-\infty}dk_1\frac {I(k_1)}{k-k_1},\; \; \; \; \; \; \nu_o=\frac {2^{1/4}\pi }{8}\frac {\Gamma (3/4)}{\Gamma (1/4)}\int \limits ^{+\infty}_{0}\frac {4\pi v^4dv\nu }{(2\pi )^{3/2}v_T^5}\exp \left[ -\frac{v^2+v_a^2}{2v_{T}^2}\right] \label {15}
\end{equation}
\section {Discussion.}
Trapping of thermal particles is essential for damping of Alfv\'en waves if the frequency of collisions is small enough. For trapped particles $|\mu |<\mu _\ast $, where $\mu _\ast \sim \delta B/B$ for Alfv\'en waves. Hence the escape time is $t_{esc}\sim \mu _\ast ^2/\nu $. It should be compared with the period of particle oscillations inside the trap $T\sim (kv_T\mu _\ast )^{-1}$. This gives us the condition for saturation of nonlinear damping:
\begin{equation}
 \nu << kv_T\left( \frac {\delta B}B\right) ^3 \label {16}
\end{equation}
The saturated damping rate can be estimated as the unsaturated damping rate multiplied by the ratio $T/t_{esc}$. It is easy to see that such an estimate is in accordance with expression (15). In our self consistent model of galactic wind flow (Zirakashvili et al., 1996, Ptuskin et al., 1997) where unsaturated damping rate was used, $\delta B/B\sim 10^{-2}$ and is determined by the power of cosmic ray sources in the Galactic disk. For this case the critical value for the collision frequency is $10^{-12}s^{-1}$ for a wavenumber $k\sim 10^{-13}cm^{-1}$ that is in resonance with 1 GeV cosmic ray protons. This value is close to the value of the collision frequency of a hot rarefied plasma with number density $10^{-3}cm^{-3}$ and temperature $10^6 K$. Therefore, in the absence of other scattering process, trapping effects might be relevant for Alfv\'en wave damping in our Galaxy.\\
Another important feature of saturated damping is the possibility of not only damping but also energy transfer to smaller wavenumbers. This property is absent for unsaturated damping of unpolarized ($I(k)=I(-k)$) waves. Such energy transfer can be important for diffusive shock acceleration because it permits small energy particles to generate Alfv\'en waves that are in resonance with particles of greater energies and, hence determines the rate of acceleration.\\
{\bf Acknowledgment}. During this work V.N.Z. and V.S.P. were supported by RFBR (98-02-16347), INTAS (95-16-711-23) and ``Astronomy'' (1.3.8.1) grants of Russian Academy of Sciences. The work of V.N.Z. was also supported by Grant for young scientists of Russian Academy of Sciences, and by SFB 328 of the Deutsche Forschungsgemeinschaft (DFG) during his visit to Max-Planck-Institut f\"ur Kernphysik in Heidelberg. V.N.Z. thanks the National Advisory Committee for financial support for attending of 26th ICRC.

%
%
%
\vspace{1ex}
\begin{center}
{\Large\bf References}
\end{center}
%
Achterberg,A. 1981, A\&A. 98,161\\
Berezinsky,V.S., Bulanov,S.V., Dogiel,V.A., Ginzburg,V.L., \& Ptuskin,V.S. 1990,Astrophysics of Cosmic rays, North-Holland Publ. Comp.\\
Chandrasekhar,S. 1960, Plasma Physics, The Univ. Chicago Press\\
Fedorenko,V.N. 1992, Astroph. Space Phys., Sov. Sci. Rev.E, ed.
R.A.Sunyaev 8,1\\
Galeev,A.A.\& Sagdeev,R.Z. 1979, Reviews of Plasma Physics 7, M.A.Leontovich (ed.), Consultants Bureau, New York\\
Kulsrud,R.M. \& Pearce,W.P. 1969, ApJ 156,445\\
Kulsrud,R.M. 1978,Astronomical papers dedicated to B.Str\o mgren, Copenhagen Univ. Obs.p.317\\
Lee,M.A. \& V\"olk,H.J. 1973, Astrophys. Space Sci. 24,31\\
Lerche,I. 1967, ApJ 147,689\\
Ptuskin,V.S., V\"{o}lk,H.J., Zirakashvili,V.N.,\& Breitschwerdt,D. 1997, A\&A
321,434 \\
V\"{o}lk, H.J.\& Cesarsky, C.J. 1982, Z. Naturforsch. 37a,809\\
Wentzel,D.G. 1969, ApJ 156,303\\
Zirakashvili,V.N., Breitschwerdt,D., Ptuskin,V.S.,\& V\"{o}lk,H.J. 1996, A\&A 311,113\\

\end{document}